\documentclass[12pt]{article}
\usepackage{graphicx}
\usepackage{aasms4}

\def\lapprox{\hbox{\lower .8ex\hbox{$\,\buildrel < \over\sim\,$}}}
\def\gapprox{\hbox{\lower .8ex\hbox{$\,\buildrel > \over\sim\,$}}}

\begin{document}

\title{The origin of the cosmic gamma--ray background in the MeV range}

\author{Pilar Ruiz--Lapuente$^{1,2,3}$, Lih-Sin The$^{4}$, Dieter H.
Hartmann$^{4}$, Marco Ajello$^{4}$, Ramon Canal$^{3,5}$, Friedrich K. 
R\"opke$^{6,7}$, Sebastian T. Ohlmann$^{6}$, Wolfgang Hillebrandt$^{2}$}

\altaffiltext{1}{Instituto de F\'{\i}sica Fundamental, Consejo Superior de 
Investigaciones Cient\'{\i}ficas, c/. Serrano 121, E--28006, Madrid, Spain}
\altaffiltext{2}{Max--Planck--Institut f\"ur Astrophysik, 
Karl--Schwarzschild--Str. 1, D--85748 Garching bei M\"unchen, Germany}
\altaffiltext{3}{Institut de Ci\`encies del Cosmos (UB--IEEC),  c/. Mart\'{\i}
i Franqu\'es 1, E--08028, Barcelona, Spain} 
\altaffiltext{4}{Department of Physics and Astronomy, Clemson University, 
SC 29634, USA}
\altaffiltext{5}{Departament d'Astronomia i Meteorologia, Universitat de
Barcelona, c/. Mart\'{\i} i Franqu\'es 1, E--08028 Barcelona, Spain}
\altaffiltext{6}{Institute of Theoretical Physics and Astrophysics, University 
of W{\"u}rzburg, D--97074, W\"urzburg, Germany}
\altaffiltext{7}{Heidelberg Institute for Theoretical Studies, 
Schloss--Wolfsbrunnerweg 35, D--69118, Heidelberg, Germany}

\begin{abstract}

There has been much debate about the origin of the diffuse $\gamma$--ray background in the MeV range. At lower energies,
 AGNs and Seyfert galaxies can explain the background, but not above $\simeq$0.3 MeV. Beyond $\sim$10 MeV blazars appear
 to account for the flux observed. That leaves an unexplained gap for which different candidates have been proposed, 
including annihilations of WIMPS. One candidate are Type Ia supernovae (SNe Ia). Early studies concluded that they were
 able to account for the $\gamma$--ray background in the gap, while later work attributed a significantly lower contribution to them.    

All those estimates were based on SN Ia explosion models which did not reflect the full 3D hydrodynamics of SNe Ia explosions. 
In addition, new measurements obtained since 2010 have provided new, direct estimates of high-z SNe Ia rates beyond $z\sim$2.
 We take into account these new advances to see the predicted contribution to the gamma--ray background.

We use here a wide variety of explosion models and a plethora of new measurements of SNe Ia rates. SNe Ia still fall short of the observed
 background. Only for a fit, which would imply $\sim$150\% systematic error in detecting SNe Ia events, do the theoretical predictions approach 
the observed fluxes. This fit is, however, at odds at the highest redshifts with recent SN Ia rates estimates. Other astrophysical sources such
 as FSRQs do match the observed flux levels in the MeV regime, while SNe Ia make up to 30--50\% of the observed flux.

\end{abstract}

\keywords{gamma rays: difuse background; supernovae; galaxies: active}

\section{Introduction}

The cosmic gamma--ray background is diffuse and its origin
is diverse and remains partly unknown at various energy ranges. On the low 
energy side, from X--ray energies up to around 0.3 MeV, AGNs and Seyfert 
galaxies provide most of the emission 
(Madau et al. 1994; Ueda et al. 2003). At $E \gapprox$ 
0.3 MeV, the spectrum of AGNs and Seyfert galaxies sharply cuts off.  
From 50 MeV to the GeV range, blazars seem to be responsible for the 
observed flux (Zdziarski 1996; Sreekumar et al. 1998). However, 
the latest results from {\it Fermi}, in the GeV range, which show a 
higher gamma--ray background at GeV energies than previous results 
from EGRET, have called into question the former attribution to blazars as 
the main source (see discussions in Lacki, Horiuchi \& Beacom 2014). 
Nevertheless, recent examinations of this issue find that at energies
 $>$ 100 MeV, 
blazars account for $\sim$50\% of the background, while the other half is
contributed by star--forming galaxies and radio galaxies (Ajello et al. 2015; 
Di Mauro \& Donato 2015). 

The measurements in the MeV range have been provided by various 
space missions. The first exploration of the region between 1  
and 5 MeV was made by the APOLLO 15/16 missions (Trombka et al. 1977). 
The reanalysis of the Apollo data, the measurements from HEAO--A4 
(Kinzer et al. 1997), the Solar Maximum Mission (Watanabe et al. 1999a), 
and COMPTEL (Kappadath et al. 1996; Weidenspointner 2000) provided the basic
empirical results on the diffuse gamma--ray background, in 
the range from 100 keV to 10 MeV. The slope of the emission spectrum 
exhibits a steep decrease with increasing energy, from a few hundred keV to 10 MeV, 
changing to a flatter slope around 10 MeV and beyond, revealing 
the need of an intense extragalactic source in the MeV window. In this 
range of energies, the discussion of the origin of the background was 
revived at the beginning of the XXIst century and continued up 
to the present time (see Lacki, Horiuchi \& Beacom 2014).

Much of the current discussion centers on the possibility 
that Type Ia supernovae are able to produce the observed flux in the 0.3---3 MeV range,
filling the gap between Seyferts and blazars. 
This possibility was first suggested by Clayton and Silk (1969), and further studied 
by The et al. (1993) and Watanabe et al. (1999b) (hereafter  W99b). At those times, there 
were no empirical rates of SNe Ia available, in particular the function $R_{Ia}(z)$, was not
measured at the required high redshifts. 
Therefore, the authors had to estimate fluxes on the basis of the better known star formation 
rates in galaxies and a convolution with an assumed delay time distribution
between the birth of the progenitor binary systems and the explosions.  
Depending on the delay time since birth of the system, the SNe Ia 
rate could be either too low or just right, to account for the measured 
level of the gamma--ray background (W99b). Ruiz--Lapuente, Cass\'e \& 
Vangioni--Flam (2001) (hereafter RCV01) performed a study of the
SNe Ia contribution in the MeV region, by considering a wide range of 
star formation rates $\dot\rho_{*}(z)$ in galaxies and different efficiencies 
in the production of SNe Ia, for each star formation rate. 
Ruiz-Lapuente (2001) found that not much under 1000 M$_{\odot}$ going 
into star formation gives rise to one SN Ia (720 $\pm$ 250 M$_{\odot}$ produce 
1 SNIa). The efficiency, defined as 
\begin{equation}
{\cal E}_{Ia}(z) = {R_{Ia}(z)\ yr^{-1} Mpc^{-3}\over \dot\rho_{*}(z)\ M_{\odot}\ 
yr^{-1} Mpc^{-3}} 
\end{equation}
was 1.41 $\pm$ 0.35 10 $^{-3}$ M$_{\odot}$$^{-1}$ $h_{65}^{2}$. In that work, 
they found that the delay time between star formation and supernova explosion
could shift the estimate up or down, but that most star formation 
histories would average out the effect, since the distribution of SNe Ia
delay times appeared to be broad. The result of this study 
was that SNe Ia yield a background emission, in the MeV range, that can
explain the extragalactic emission measured by COMPTEL and SMM. 

Just shortly afterwards, the topic was addressed by other authors 
(Iwabuchi \& Kumagai 2001;  Ahn, Komatsu \& H\"oflich 2005 ---hereafter 
AKH05---; Strigari et al. 2005), and was more recently considered again 
by Horiuchi \& Beacom (2010, hereafter HB2010). The latter authors obtained a 
range of possibilities concerning
the SN Ia contribution, but found that they are in general not able  
to account for the MeV background. The earlier "negative" results can be traced
to the first high--$z$ SNe Ia rates obtained by Dahlen et al. (2004, hereafter
D04) which peak at a smaller $z$ than current rates (Okumura et al. 2014; 
Rodney et al. 2014; Graur et al. 2014). However, HB2010
do not use the Dahlen et al. (2004) rates, but a common fit to data availablle 
up to 2010. These rates will be compared with most recent measurements.

Both W99b and RCV01 (see also similar findings by The et al. 1993 
and by Zdziarski 1996) had concluded that there was room for SNe Ia to account 
for the unexplained MeV background. They did so by integrating the full gamma-ray 
spectrum emerging from SN Ia models during the early period of significant production 
by radioactive decay. In this context one includes the production of gamma-ray photons 
as well as their transport in the expanding ejecta. AKH05 take a SNIa rate (from D04) that 
is one order of magnitude smaller than the one adopted by W99b. This may explain most 
of the discrepancy in the results. 
 HB2010 
calculations take the continuum plus line $\gamma$--ray 
spectra from the W7fm model (fully mixed W7 model) 
shown in W99b and the 5p0z22.23 model shown 
in  AKH05 (2005) (private communication). We find  agreement
with their results when using the same SNe Ia rates as 
HB2010 did.
  
One interesting feature of the SNe Ia contribution, which is shared by the 
 calculations presented in W99b, RCV01, AKH05, HB2010 is that the predicted SNe Ia
gamma--ray background contribution has a spectral shape (power law in energy)
that runs parallel to the observed fluxes up to 3 MeV. This match in slope is one of the 
primary reasons for considering SN Ia as a significant contributor, and to focus on the 
disagreement with the flux normalization.

Due to the recent appreciation of SN Ia as cosmological probes, a significant increase 
in dedicated observational programs produced a greatly improved knowledge of their rates, 
for a wide range of redshifts (some of the most recent contributions were mentioned 
above).
We also know well the cosmological parameters required for the calculation 
 ($H_{0}$, $\Omega_{M}$, $\Omega_{\Lambda}$). Moreover, for the 
first time, a nearby SN Ia (SN 2014J) has been detected in the MeV range (with INTEGRAL: 
Churazov et al. 2014; Diehl et al. 2014). The latter authors account for 
the gamma--ray data with a white dwarf explosion having a small amount of 
$^{56}$Ni at the outskirts and around 0.5 $M_{\odot}$ in the innermost core. 
This model is similar to those used in the previous calculations of the 
gamma--ray background (and also to those that will be used here). 
Thus, we now have a firmed up grasp of the gamma--ray flux emitted by SNe Ia.

These developments led us to once more calculate the gamma--ray background
from SNe Ia, now based on improved knowledge of the theoretical ingredients. 
This paper is organized as follows. In Section 2, we discuss the different numerical 
methods used to compute the gamma--ray fluxes from given SNe Ia models. 
The models used for the present study are described in Section 3.  
In Section 4, we present an update of the SNe Ia rates at various $z$ and 
show the interpolations used for the background calculations. In Section 5, we 
explain the gamma--ray background formulation and present the results.
Section 6 discusses the results and the final Section contains our conclusions.

\section{Gamma--ray escape}

In order to derive the emergent spectrum predicted by supernova models, we
use the Monte Carlo code described in Burrows \& The (1990) and 
The, Burrows \& Bussard (1990), modified to include Bremsstrahlung X--ray
production and the iron fluorescence line at $\sim$6.4 keV (Clayton \& 
The 1991; The, Bridgman \& Clayton 1994). This code follows the
gamma--ray emission of radioactive nuclei such as $^{44}$Ti, $^{56}$Ni, 
$^{56}$Co, $^{57}$Co. The code was used in the predictions by W99b. 

In RCV2001, a modified version of the Ambwani \& Sutherland
(1988) code was used, which was also employed by Lehoucq, Cass\'e \& 
Cesarsky (1989) to study the radioactive output of SN1987A. It was further 
modified (Ruiz-Lapuente et al. 1993) to include positronium formation 
in the $^{56}$Co decay, 
giving rise to two--photon (parapositronium) and three--photon decay 
(orthopositronium), the energy distribution being that derived by Bussard, 
Ramaty \& Drachman (1979). The Monte Carlo routine was then used for the 
study of gamma--rays from SNe Ia (Ruiz--Lapuente et al. 1993). A comparison 
of the predicted gamma--ray emission, for model W7 (Nomoto, Thielemann \& 
Yokoi 1984), with that from W99b shows the gamma--ray fluxes predicted from 
the code used in RCV2001 to be 10\% higher in the MeV regime than those  
from W99b. Differences of the gamma-ray yields resulting from different gamma--ray 
transport codes were studied in detail by Milne et al. (2004), who show that variations 
of $\sim$10\% can arise from the different physics and methods employed. 
With the use of the same code as in W99b, Burrows \& The (1990), and The, Burrows \& 
Bussard (1990), we are on the  conservative side concerning the predicted gamma--ray 
emission (see Milne et al. 2004).

\section{Input models}

A wider range of predicted emerging fluxes in the MeV domain is obtained when using 
different input models for the SNe Ia, as recently shown by 
The \& Burrows (2014). Looking at these results, one notices that there are 
$\sim$20\% variations, in the MeV range, between various input models. Here
we pick five different models in an attempt to delineate the upper limit 
constraint on the gamma--ray emision from normal SNe Ia. These models are 
W7 (Nomoto, Thielemann \& Yokoi 1984), the fully mixed W7 model (W7fm) used in 
W99b (both models synthesize 0.58 M$_{\odot}$ of $^{56}$Ni), the W7dt model of 
Yamaoka et al. (1992) (this one predicts a larger escaping
gamma--ray flux than most of the other models explored by The \& Burrows 
2014 as it synthesizes  0.78 M$_{\odot}$ of $^{56}$Ni, a quantity
slightly larger than in a normal SNIa),
 the spherically--averaged 3D model N100 of a delayed detonation 
(R\"opke et al. 2012; Seitenzahl et al. 2013), and the (also averaged) 3D 
model of the violent merging of two WDs of Pakmor et al. (2012).
 Those two 3D averaged models synthesize, respectively, 0.604 M$_{\odot}$ of
$^{56}$Ni in the model N100 and 0.62 M$_{\odot}$ in the averaged 3D violent 
merging of two WDs, quantities within the canonical range of 0.55-0.65 M$_{\odot}$
of $^{56}$Ni for normal SNe Ia.

The first three models have been widely discussed in the literature.
Their gamma--ray emissions are compared in Figure 1.
The 3D model N100 of R\"opke et al. (2012) corresponds to an initially
isothermal WD, made of C and O in equal parts, with a central density
of $2.9 \times 10^{9}$ g cm$^{-3}$. It was ignited in 100 sparks (hence its
name), placed randomly in a Gaussian distribution within a radius of 150 km
from the WD's center. After an initial deflagration phase, a detonation was
triggered at every location of the flame where the criterion for a
deflagration--to--detonation transition of Ciaraldi--Schoolmann et al. (2013) 
was met. 

The model from Pakmor et al. (2012), also a 3D model, simulates the violent 
merger of a 1.1 M$_{\odot}$ and a 0.9 M$_{\odot}$ WD. 
As the material of the tidally disrupted secondary WD hits the primary WD, a 
hot spot forms which leads to the ignition of C burning. The conditions 
reached there are sufficient to trigger a detonation that burns the merged 
object and leads to the explosion.

The spectral evolution of the gamma--ray emission from the two 3D models 
above is shown in Fig. 1 of Summa et al. (2013), together with the spread 
due to different viewing angles for the maximum--light epochs in gamma rays
of both models. Here, however, we use the spherically--averaged models, with the
same treatment of gamma--ray escape as for the three 1D models (W7, 
W7fm, and W7dt). Their comparison with W7fm is shown in Figure 2. 

Ruiter et al. (2013) have shown that sub--Chandrasekhar pure detonation 
models can reproduce the 
observed peak--magnitude distribution of SN Ia (Li et al. 2011). 
The brightness of the explosion is mainly determined by the $^{56}$Ni 
mass synthesized in the detonation of the primary WD. Sim et al. (2010)  
derived a relationship between the mass of that WD and the expected peak 
bolometric brightness. Ruiter et al. (2013) use the population 
synthesis data from Ruiter et al. (2011) to derive the theoretical 
peak--magnitude distribution. 

Here we estimate the contribution of this channel to the gamma--ray 
background, based on the fact that the distributions of WD and $^{56}$Ni 
masses peak at 1.1 M$_{\odot}$ and 0.6 M$_{\odot}$, respectively. In 
Figure 3 we compare the gamma--ray spectrum of a representative model of 
such mergings with that of model W7fm. The major difference between models
is in the X--ray domain between 1 and 50 keV. However, in the range of MeV,
the difference between these models that reproduce the bulk of SNe Ia 
in the optical is of order 5$\%$. 
Fig.4 shows the difference between the models in different energy ranges.

What has been shown in this Section is that the background arising from 
different explosion models, associated with different progenitor origin channels (but 
all aiming at explaining normal SNe Ia) induces only a small variance.  
The convolution of the gamma--ray yield resulting from several models, here exemplified 
by model W7fm, with the supernova rates is explored in Section 5.

\begin{figure}
\centering
\includegraphics[width=0.6\columnwidth]{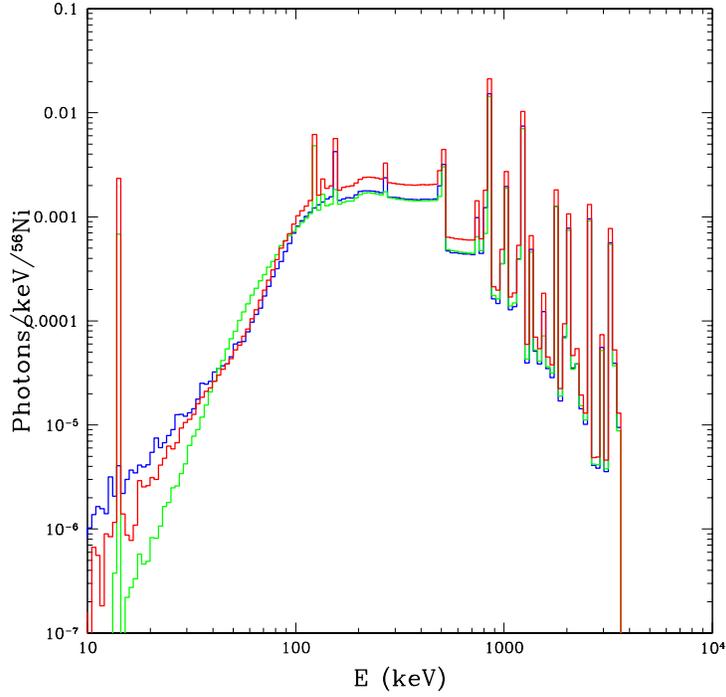}
\caption{The total X--ray continuum and gamma--ray line fluxes per 
$^{56}$Ni nuclei, integrated over the first 600 days. The green line shows the 
corresponding fluxes from the calculations of The et al. (1993) for the W7 
model (Nomoto, Thielemann, \& Yokoi 1984), the blue line shows the fluxes 
for the fully mixed W7 model (W7fm) used in W99b, and the red line shows 
the fluxes for the W7dt model from Yamaoka et al (1992),
(see also The \& Burrows 2014).} 
\label{Figure 1}
\end{figure}

\begin{figure}
\centering
\includegraphics[width=0.6\columnwidth]{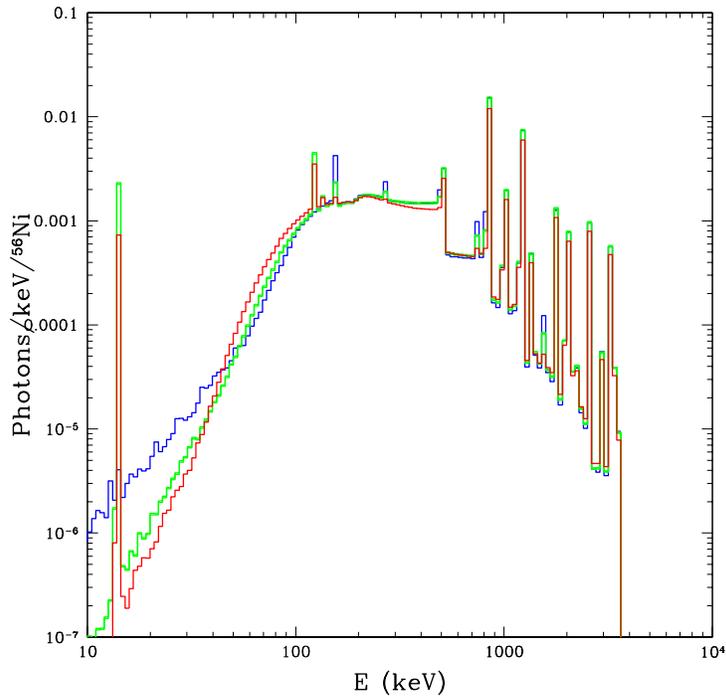}
\caption{The total X--ray continuum and gamma--ray line fluxes 
per $^{56}$Ni nuclei, integrated over the first 600 days comparing the calculations for 
the fully mixed W7 model (W7fm) used in W99b (blue line), for the (spherically 
averaged) delayed--detonation 3D model N100 of R\"opke et al. (2012) (green 
line), and for the (also averaged) 3D model of the violent merging of two WDs 
of Pakmor et al. (2012) (red line).} 
\label{Figure 2}
\end{figure}

\begin{figure}
\centering
\includegraphics[width=0.6\columnwidth]{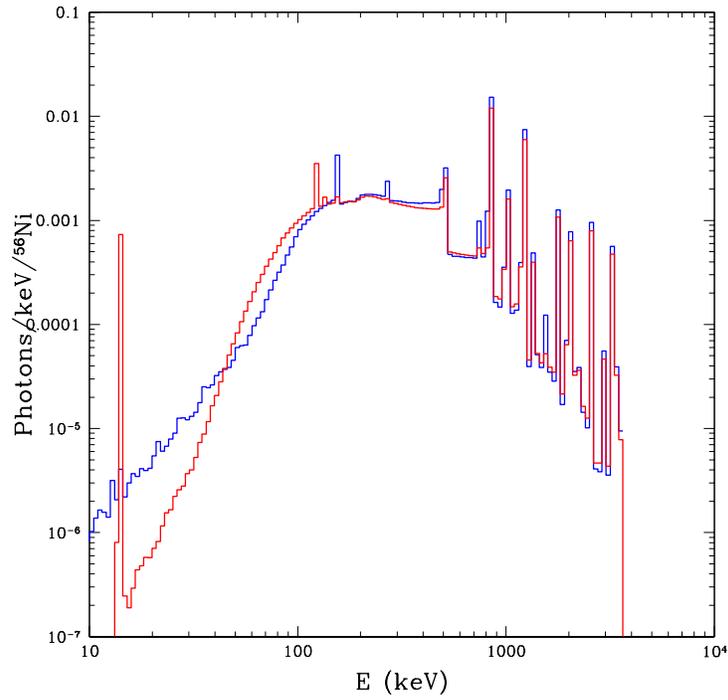}
\caption{Total time-integrated X--ray continuum and gamma--ray line fluxes 
per $^{56}$Ni nucleus , comparing the calculations for an average model of 
the sub--Chandrasekhar pure detonation models (Sim et al. 2010; Ruiter et al. 
2013) (red line), and for the W7fm model of W99b (blue line).} 
\label{Figure 3}
\end{figure}

\begin{figure}
\centering
\includegraphics[width=0.6\columnwidth]{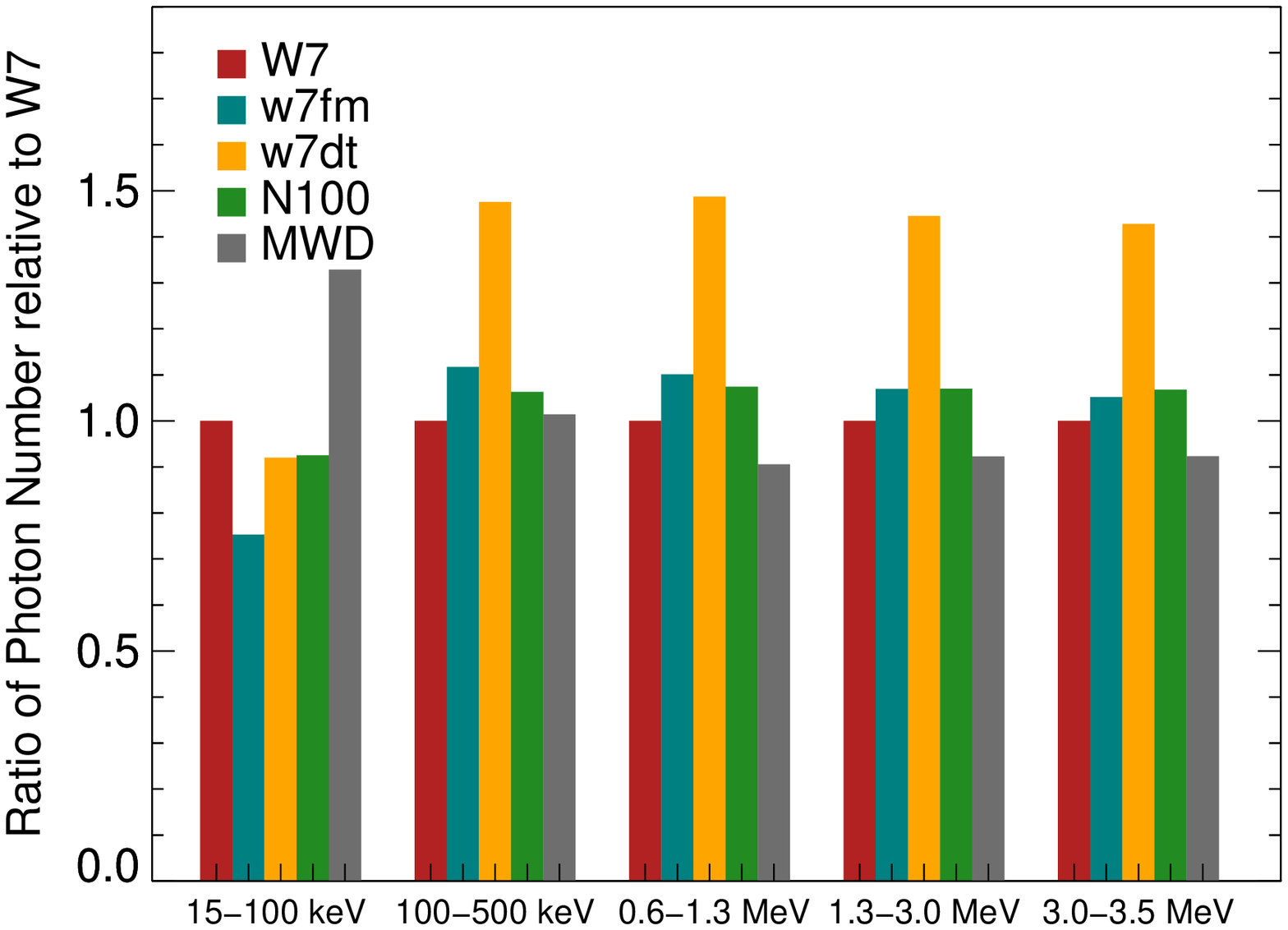}
\caption{The ratio of number of emergent photons relative to model W7. The bar graph shows the ratio of integrated number of emergent photons of models considered here relative to W7.  Note that photons in energy bands below ~0.6 MeV are not relevant to the CGB near MeV region, which is our main consideration, but we plot them for completeness. Please note the good agreement within 5 $\%$ 
of the 3D models with the w7fm model as well as with the W7 model. The model 
w7dt has more $^{56}$Ni mass than a normal SNIa. }
\label{Figure 4}
\end{figure}

\section{SNe Ia rates}

Several groups have obtained SNe Ia event rates using either 
ground--based facilities, the combination of ground--based and space facility 
such as the Subaru/XMM--Newton Deep Survey (Okumura 
et al. 2014 being the most recently published search of this kind), 
or the {\it Hubble Space Telescope} (Rodney et al. 
2014, and Graur et al. 2014 have presented the most recent results from 
such searches). One goal of those surveys is to gauge the 
delay--time distribution (DTD) of SNe Ia, i.e., the time that the SNe Ia 
progenitors spend between formation and explosion. 

 Figure 5 shows how the different fits to compilations by 
Okumura et al. (2014), Graur et al. (2014) and Rodney et al. (2014)
pass through the data up to z $\sim$ 1 with similar lines, but 
they differ significantly at larger redshifts. For a close look at low and high $z$, we 
plot in Figure 6  all measurements made up to now, taken from the compilation
 in Table 4 of Graur et al. (2014) (superseded ones excluded)\footnote{In order of 
increasing redshuft, they are from: Cappellaro et al. (1999), Li et al. (2011a), 
Dilday et al. (2010), Dilday et al. (2008), Dilday et al. (2010), Graur \& 
Maoz (2013), Blanc et al. (2004), Hardin et al. (2000), Dilday et al. (2010), 
Rodney \& Tonry (2010), Perrett et al. (2012), Horesh et al. (2008), Dilday et 
al. (2010), Dilday et al. (2010), Perrett et al. (2012), Botticella et al. 
(2008), Dilday et al. (2010), Rodney \& Tonry (2010), Perrett et al. (2012), 
Graur et al. (2014), Rodney \& Tonry (2010), Perrett et al. (2012), Graur et 
al. (2014), Rodney \& Tonry (2010), Perrett et al. (2012), Tonry et al. 
(2003), Dahlen et al. (2008), Pain et al. (2002), Rodney \& Tonry (2010), 
Perrett et al. (2012), Melinder et al. (2012), Rodney \& Tonry (2010), 
Perrett et al. (2012), Graur et al. (2011), Rodney \& Tonry (2010), Perrett 
et al. (2012), Barbary et al. (2012), Dahlen et al. (2008), Rodney \& Tonry 
(2010), Perrett et al. (2012), Graur et al. (2014), Rodney \& Tonry (2010), 
Perrett et al. (2012), Rodney \& Tonry (2010), Perrett et al. (2012), Barbary 
et al. (2012), Dahlen et al. (2008), Graur et al. (2011), Barbary et al. 
(2012), Graur et al. (2014), Dahlen et al. (2008), Graur et al. (2014), 
Dahlen et al. (2008), and Graur et al. (2011), and are shown as open
circles in the Figure.}

The estimates of the SNe Ia rates are derived in different ways.
Okumura et al. (2014), following previous work by the {\it Supernova 
Cosmology Project}, select objects having a flux 
larger than 5\,$\sigma_{b}$, where $\sigma_{b}$ is the background 
fluctuation within an aperture of 2 arcseconds in diameter. These objects 
are classified as transients. The authors require that objects show at least a 
5\,$\sigma_{b}$ increase in 2 or more epochs, in the $i'$--band. This 
requirement is accounted for in the rate calculation when computing the 
control time. For the supernova hosts for which spectroscopic redshifts 
are not available, they use photometric redshifts from the probability 
distribution function of the host galaxy redshifts. They discriminate 
against AGN using X--ray data from observations with XMM-Newton. 
Template light curves help to discriminate SNe Ia from  SNe Ib/II. 

The final count of SNe Ia detected leads to the number of supernovae  
per comoving volume unit, $r_{V}(z)$. The number of SNe Ia expected 
in a redshift bin $(z_{1} < z < z_{2})$ is given by (Okumura et al. 2014): 
\begin{equation}
N_{exp}(z_{1} < z < z_{2}) = \int_{z_{1}}^{z_{2}}{r_{V}(z){CT(z)\over 1 + z} 
{\Theta\over 4\pi}V(z)}dz
\end{equation}
where $V(z)dz$ is the comoving volume in a redshift slice of thickness $dz$, 
$\Theta$ is the solid angle observed in the survey (in units of steradians),
and $CT(z)$ is the observer frame ``control time'', i.e. the total time
for which the survey is sensitive to a SN Ia at redshift $z$. The SNe Ia 
rates per comoving volume of the Subaru/XMM--Newton Deep Survey (Okumura
et al. 2014) are represented by blue triangles in Figure 5 and 6 together with
the compilation of rates at all different z.  They include a 
generous 50 \% error in the systematic  uncertainties due to dust
obscuration leading to lack of detection of SNe Ia. In the figures, 
we also plot the 1$\sigma$ upper and lower bounds of Okumura et al.(2014)
rates as two blue long--short dashed lines. They represent the total, statistical and 
systematic, uncertainties  of the points measured by Okumura et al. (2014). 

Graur et al. (2014) presents a compilation of all previous measurements
of SNe Ia rates and new rates obtained by the CLASH collaboration in the
range 1.8 $<$ z$<$ 2.4. The procedure they use to compute the rates
has aspects in common with Okumura et al. (2014) approach, as 
they look for candidates
in this case with a flux 3$\sigma$ above the background and they rely on a 
good knowledge of the redshift of either the SN or its host galaxy. But 
it also includes some Bayesian approach in the classification of the
SN with a probability obtained by comparing the light curve of the 
supernova with light curve models, as in Rodney et al. (2014).

Rodney et al. (2014) (for the CANDELS collaboration) use technique and requirements 
different from those of Okumura et al. (2014). They use a  Bayesian photometric 
classification algorithm  based on the light curves of high--redshift 
supernovae. The main uncertainties in this approach are: the redshift--dependent
prior describing the relative fraction of SN that are SNe Ia, and
the assumed  distribution of dust extinction  values. The host--based prior 
comes from the algorithm {\it galsnid} of Foley and Mandel (2013), which
exploits the relationships between SN and their host environments. 
They adopt  a distribution of host galaxy extinctions for the dust extinction
that then enters as a further probability into the SNe Ia classification
algorithm. The mid--rate plus mid--dust combination gives their baseline 
classification probability, which determines how much each individual SN 
contributes to the total count of observed SNe Ia. The upper and lower bounds 
make the error bars. The SNe Ia rates from the CANDELS collaboration are 
plotted as green squares in Figures 5 and 6. 

The utilization of informed Bayesian priors to estimate the SNe Ia rate as 
a function of redshift was published earlier by Graur et al. (2001). 
Kuznetsova et al. (2007) had followed a Bayesian approach as well, 
to determine the rates in four redshift bins.

The most conservative SNe Ia rate error bars derived up to date are those from
Barbary et al.(2012). We show them as inverted black triangles in Figure 5 and 6. 
The study considers different dust models in estimating SNe Ia rates where it 
gives the highest estimates of the observed SNe Ia rates, and also the largest 
uncertainties. The large error bars also result from  the large uncertainties of 
dust extinction estimates. 

The D04 data are shown as red circles in Figure 5. The fit to these data 
appears as the red long--dashed line in the Figure. As it can be seen, we have 
fitted the  D04 data in a similar way as in AKH05. We have considered, as in
that work, that the first errorbar is reduced by taking the mean value  
of the upper and lower error bars of the first $z$.

 One controversial aspect in the 1$\sigma$ and 2$\sigma$ error bars 
from Okumura et al. (2014) is the way they are calculated. If we were to
take the 1$\sigma$ upper and lower
 limits to the SN Ia rates by propagating the errors in the fitting 
parameters given by their expressions (10)--(12), the upper  1$\sigma$
limit would be at odds with the data from all other surveys, and 
also well beyond that indicated by the uncertainties of their own data. 
There is no questionable deduction, however,  of the 1$\sigma$ errors of their 
individual measurements presented for first time in their 2014 paper.
Nevertheless, the extrapolation of the fitting formula beyond their data 
departs significantly from what has been found by other authors 
(Rodney et al. 2014; Graur et al. 2014). We keep their fitting formula, 
given that discrepancies start beyond the redshift range from within which 
most of the background  arises, and we use the 1$\sigma$ and 2$\sigma$ 
uncertainties as derived from their data points.

To date there remain divergences in the measured SN Ia rates, 
despite  the considerable increase of the data base in recent years. They
reflect the differences in the procedures followed to derive the rates from
the observations. In particular, how the SNe Ia rates estimate accounts 
for dust obscuration.

The approach by Okumura et al. (2014) is well tested. It clearly
discards SNe Ia from AGNs and SNe of other types. Contamination by 
SNe Ibc is estimated and taken into account in the systematic uncertainty.
Thes authors acknowldege the significant uncertainty related to dust extinction (50 \%). 
{Note the present disparity in the treatment of dust obscuration and  the very
different approaches taken by different authors}.  
 
We just report the published rates and their corresponding gamma-ray distributions 
based on three different estimates of these rates at the highest available z to
show the level of uncertainty at high z while they are consistent at low z, where 
these three major estimates take into account all previously measured SNe Ia rates
(see Figures 5 and 6).
We calculate the gamma--ray background contribution corresponding to those SNe Ia rates.

\begin{figure}
\centering
\includegraphics[width=0.5\columnwidth]{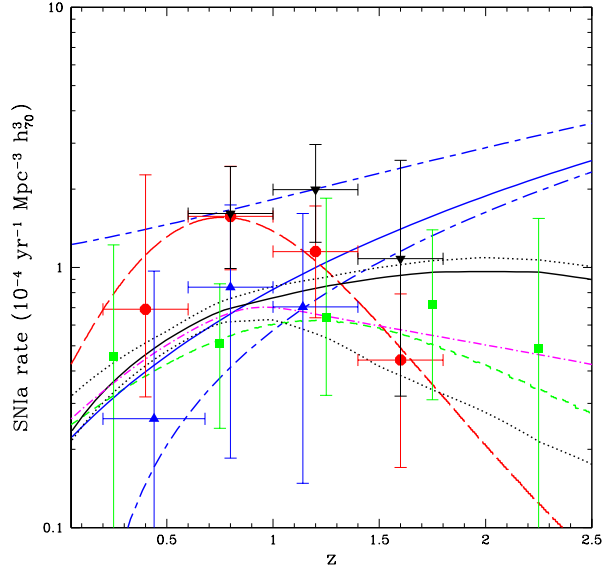}
\caption{                        
SN Ia rates measured at different redshifts and diverse fits to them. Red 
circles are D04 data from Dahlen et al. (2004); green squares come from the 
{\it CANDELS} program (Rodney et al. 2014); blue triangles are from the 
{\it Subaru/XMM-Newton Deep Survey} (Okumura et al. 2014); inverted black 
triangles correspond to the {\it HST Cluster Supernova Survey} (Barbary et al. 
2012). The red long--dashed line is a fit to the D04 data, done as in AKH05. 
The green, short-dashed line, is the same fit to the {\it CANDELS} data shown 
in Fig. 11 of Rodney et al. (2014), while the magenta dot--dashed line 
corresponds to the rates used by HB2010  (private 
communication). The solid black line is a fit to the rates given in Table 6 
of Graur et al. (2014) (see next Figure for details), and the two black dotted
lines are its $\pm1\sigma$ limits. Finally, the blue solid line is the same 
fit as in Fig. 11 of Okumura et al. (2014) to the {\it Subaru/XMM-Newton Deep 
Survey} data, and the two blue long--short dashed lines its $\pm1\sigma$ 
limits. Note that the upper 1$\sigma$ limit just fits the two highest measured
rates.  The $\pm1\sigma$ regions around the rates from Okumura et 
al. (2014) have been traced by calculating the error using the statistical and 
systematic errors of the points measured by these authors. Note that the upper
systematic errors include 50$\%$ extinction by dust, which explains why the 
upper error line is more distant from the main fit than the lower one.
}
\label{Figure 5}
\end{figure}

\vfill\eject

\begin{figure}
\centering
\includegraphics[width=0.5\columnwidth]{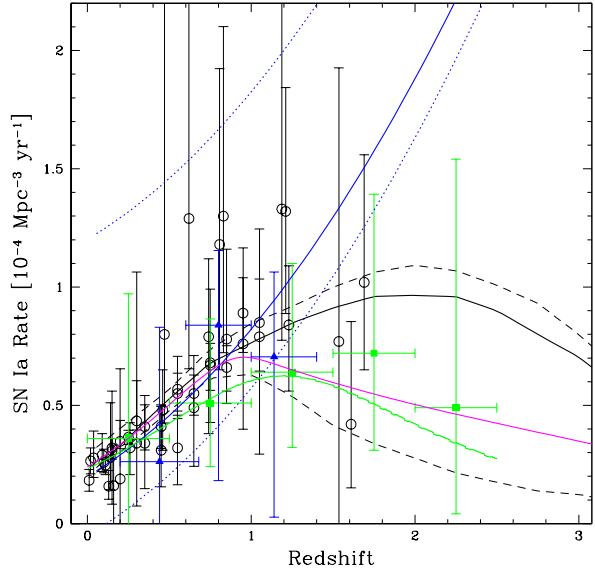}
\caption{\footnotesize{
 Different fits to the currently available SN Ia rates data, up to a 
redshift 
$z \simeq  2.25$.  The solid black line is a fit to the rates given in Table 6 
by Graur et al. (2014) 
and the two black dashed lines its $\pm1\sigma$ limits. The blue
solid line running up to the upper edge of the diagram is the fit by Okumura 
et al. (2014) to the {\it Subaru/XMM-Newton Deep Survey} data, and the two 
dotted lines the +-1$\sigma$ limits. The green solid line is the fit to the
{\it CANDELS} data (shown as green squares) alone, from Rodney et al. (2014), 
while the magenta solid line corresponds to the SN Ia rates adopted by 
HB2010 (private communication). The open circles are the 
data from different surveys, covering from $z = 0$ up to $z \simeq 1.7$ (see
text for the references). Note, when comparing with the previous Figure,  
the vertical scale is here linear instead of semilogarithmic.
The $\pm1\sigma$ regions around the rates from Okumura et 
al. (2014) have been traced by calculating the error using the statistical and 
systematic errors of the points measured by these authors. Note that the upper
systematic errors include 50\% extinction by dust, which explains why the 
upper error line is more distant from the main fit than the lower one.} 
}
\label{Figure 6}
\end{figure}

\section{Background contributions}

Based on the gamma--ray spectra in Figures 1--3 (number of escaping gamma 
photons per keV and per nucleus of $^{56}$Ni synthesized in the explosion, 
as a function of energy $E$), calculated for different SN Ia models, we 
compute the average luminosity $l_{\gamma}(E)$ (photons s$^{-1}$ keV$^{-1}$) 
of a SN Ia from the amount of $^{56}$Ni in the model and the time along 
which the SN has been emitting. In the input spectra, all the photons 
emitted from the time of the explosion to 600 days later were collected.  
Therefore, the number of active SNe Ia per unit of comoving volume, at 
any time, is that of those produced during the preceding time interval: 
$R'_{Ia}(z) = const. \times R_{Ia}(z)$, the latter being the comoving
SN Ia rate (SN yr$^{-1}$ Mpc$^{-3}$). Thus, the contribution to the 
gamma--ray background of the shell at comoving radius $r$ and with 
thickness $dr$ would be:
\begin{equation}      
dL_{\gamma}(E,z) = 4\pi R'_{Ia}(z)l_{\gamma}(E)dV(z)
\end{equation}
where
\begin{equation}
dV(z) = d_{M}^{2}(z)d(d_{M})
\end{equation}
$d_{M}$ being the proper motion distance (in Mpc). The flux received from 
that shell (in photons cm$^{-2}$ s$^{-1}$ keV$^{-1}$) will be:
\begin{equation}
dF_{\gamma}(E,z) = {1\over 4\pi d_{L}(z)^{2}} dL_{\gamma}[(z+1)E,z] 
\end{equation}
$d_{L}$ being the luminosity distance (in cm). The factor $(z+1)$, multiplying 
$E$, accounts for the redshift of the photons. We thus have:
\begin{equation}
dF_{\gamma}(E,z) = R'_{Ia}(z)l_{\gamma}[(z+1)E]{d_{M}^{2}(z)\over d_{L}^{2}(z)}
d(d_{M})
\end{equation}
Due to time dilation, there should be a factor $(1+z)^{-1}$ multiplying the 
comoving SN rate, but it is cancelled by the $(1+z)$ factor accounting for
compression of the energy bins. Then, since $d_{L} = (1+z)d_{M}$: 
\begin{equation}
dF_{\gamma}(E,z) = {1\over (1+z)^{2}}R'_{Ia}(z)l_{\gamma}[(z+1)E]d(d_{M})
\end{equation}  
$d(d_{M})$ depends on the cosmological parameters
$H_{0}$, $\Omega_{M}$, and $\Omega_{\Lambda}$, and
\begin{equation}
F_{\gamma}(E) = {c\over H_{0}}\int_{0}^{z_{lim}}{{1\over (1+z)^{2}}R'_{Ia}(z)
l_{\gamma}[(z+1)E] e(z,\Omega_{M},\Omega_{\Lambda})} dz 
\end{equation}
 We adopt $H_{0} = 70$ km s$^{-1}$ Mpc$^{-1}$, $\Omega_{M} = 0.3$, 
and $\Omega_{\Lambda}$ = 0.7 in calculating our cosmic gamma--ray background.  
Indeed, the adopted values for $\Omega_{M}$ and $\Omega_{\Lambda}$ are the 
values favored by the {\it PLANCK} collaboration (Ade et al. 2013). The  
$H_{0} = 70$ km s$^{-1}$ Mpc$^{-1}$ value is a mean value coming from the 
discussions of the various approaches to determine  $H_{0}$. 

Thus, the universe is flat and:
\begin{equation}
e(z,\Omega_{M},\Omega_{\Lambda}) = [(1+z)^{2}(1+\Omega_{M}z) - 
z(2+z)\Omega_{\Lambda}]^{-1/2}
\end{equation}
We take $z_{lim}$ = 2.5, as our various test calculations show that the 
contribution to the background from SNe Ia at higher redshifts is negligible. 
That can readily be understood because SNe Ia rates drop significantly at 
redshifts $z$ larger than $\sim$ 2. Furthermore, in order to compare 
calculated fluxes with observed values, we need  to divide the $F_{\gamma}(E)$ 
above by $4\pi$, to convert to the units of observed fluxes  
(photons cm$^{-2}$ s$^{-1}$ keV$^{-1}$ sterad$^{-1}$). 

In Figure 7, we first show the calculated cosmic gamma--ray background 
from SNe Ia model W7fm (W99b), for the SNe Ia rates of the Okumura et
al. (2014). The rate is the  main fit to the data of Okumura et al. (2014)  
at different redshifts $z$ (the blue solid line in Figures 5,6). We show 
the predicted backgrounds when considering the contributions from SNe Ia up 
to $z$ = 0.50, 1.00, 1.50, 2.00, 2.25, and 2.50, 
respectively (the increasing contributions correspond to the also increasing 
values of $z$). It can be seen that, indeed, after $z = 2$, further 
contributions to the background are negligible. In the Figure, the black squares
correspond to the {\it COMPTEL} data, analyzed by Kappadath et al. (1996), 
while the continuous line gives the results from the {\it Solar Maximum 
Mission} (Watanabe et al. 1999a), the dotted lines being the 1\,$\sigma$ upper
and lower limits.  

In order to make a consistency check, we calculate the gamma--ray background 
from rates resulting from the fit to the D04 data (red dashed line in Figure 5
and Figure 6) 
and its 1\,$\sigma$ and 2\,$\sigma$ predictions, using the W7fm model. Though 
the model is not the same as that used in AKH2005, we expect that our predictions 
should not significantly differ from theirs. Indeed, our resulting cosmic
gamma--ray background, shown in Figure 8 as a black solid line, coincides with 
that of AKH05. In Figure 8, we also plot the 1$\sigma$ upper limit in blue
 dashed line and
the 2$\sigma$ upper limit of the
cosmic gamma--ray background using the D04 SNe Ia rate as the red dashed line
confirming AKH05's conclusion that SNe Ia can not account for the observed 
gamma--ray background. However, that conclusion is mostly a direct consquence 
of 
the D04 SNe Ia rate used in the calculations, as we shall see below.  

\begin{figure}
\centering
\includegraphics[width=0.6\columnwidth]{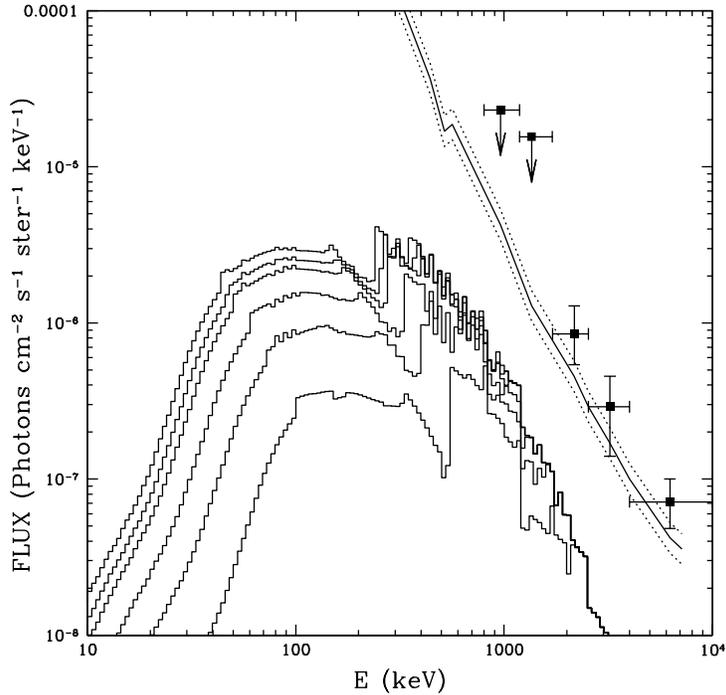}
\caption{Contributions to the gamma--ray background when considering
the mean value fit to the SNe Ia rates from Subaru--XMM (Okamura et 
al. (2014), for redshifts up to $z$ = 0.50, 1.00, 1.50, 2.00, 
2.25 and 2.50 (curves in the order of increasing backgrounds at energy $E$ 
around 100 keV correspond to the also increasing values of $z$), for model 
W7fm. One can see that, after $z$ = 2, the additional contribution becomes very small.
 Here and in the next three Figures, the black squares
correspond to the {\it COMPTEL} data, analyzed by Kappadath et al. (1996), 
while the continuous line gives the results from the {\it Solar Maximum 
Mission} (Watanabe et al. 1999a), the dotted lines being the 1\,$\sigma$ upper
and lower limits.}
\label{Figure 7}
\end{figure}

\begin{figure}
\centering
\includegraphics[width=0.6\columnwidth]{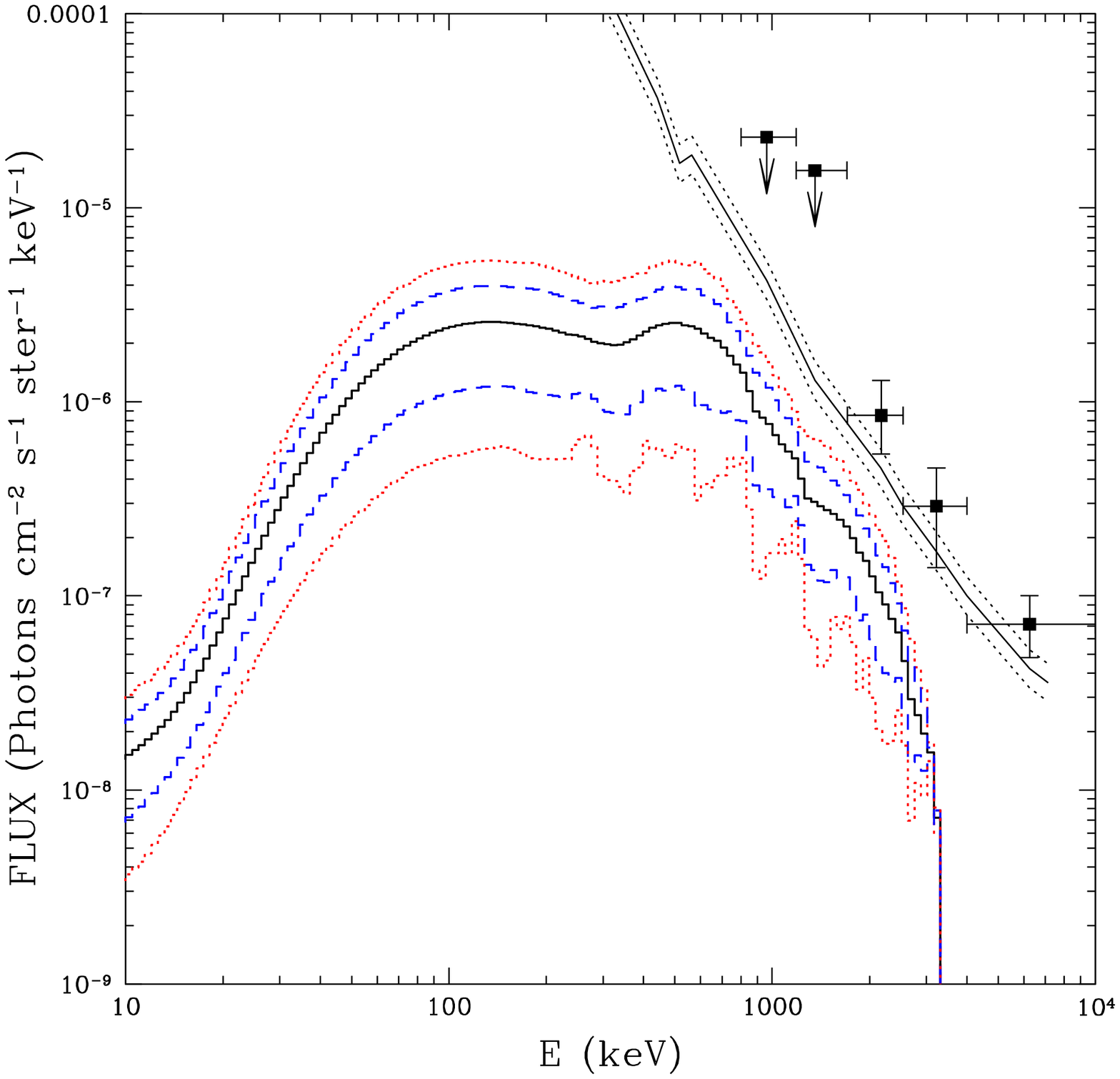}
\caption{This figure shows the gamma--ray background obtained using the mean 
fit from D04 to the SN Ia rates (see Figure 4) as the black solid lines and
the 1$\sigma$ upper and lower bounds are shown as blue dashes lines. The 
2$\sigma$ upper and lower bounds are shown  as the red dashed lines.
With those rates, the SNe Ia can not account for the observed gamma--ray 
background. We have used here the fully mixed W7 model (W7fm), as in all the 
other Figures.
 Indeed, our results for the cosmic gamma--ray background shown here as 
the black solid line coincides with that of AKH05.}
\label{Figure 8}
\end{figure}

Having calculated the gamma--ray background using the fit of the SNe Ia rate to 
the D04 data, we now consider the fit of Okumura et al. (2014). In Figure 9, 
we show the background spectrum for the mean value of the fit by Okumura et 
al. (the blue solid line in Figure 5 and 6) as the black continuous line. 
Clearly the best fit to the SNe Ia rate from the Subaru/XMM--Newton Deep 
Survey does not produce good agreement with the observed cosmic gamma--ray 
background from SMM and COMPTEL (Figure 9). In Figure 9, we also plot the background spectrum for the
+- 1$\sigma$ limits of Okumura et al. (2014) shown in Figures 5 and
 6 and their
correspondings $\pm 2\sigma$ bounds. We include the corresponding 
predictions in Fig. 9 for the gamma--ray background as they represent
the most favorable case for SNe Ia as sources in the MeV range.

\begin{figure}
\centering
\includegraphics[width=0.6\columnwidth]{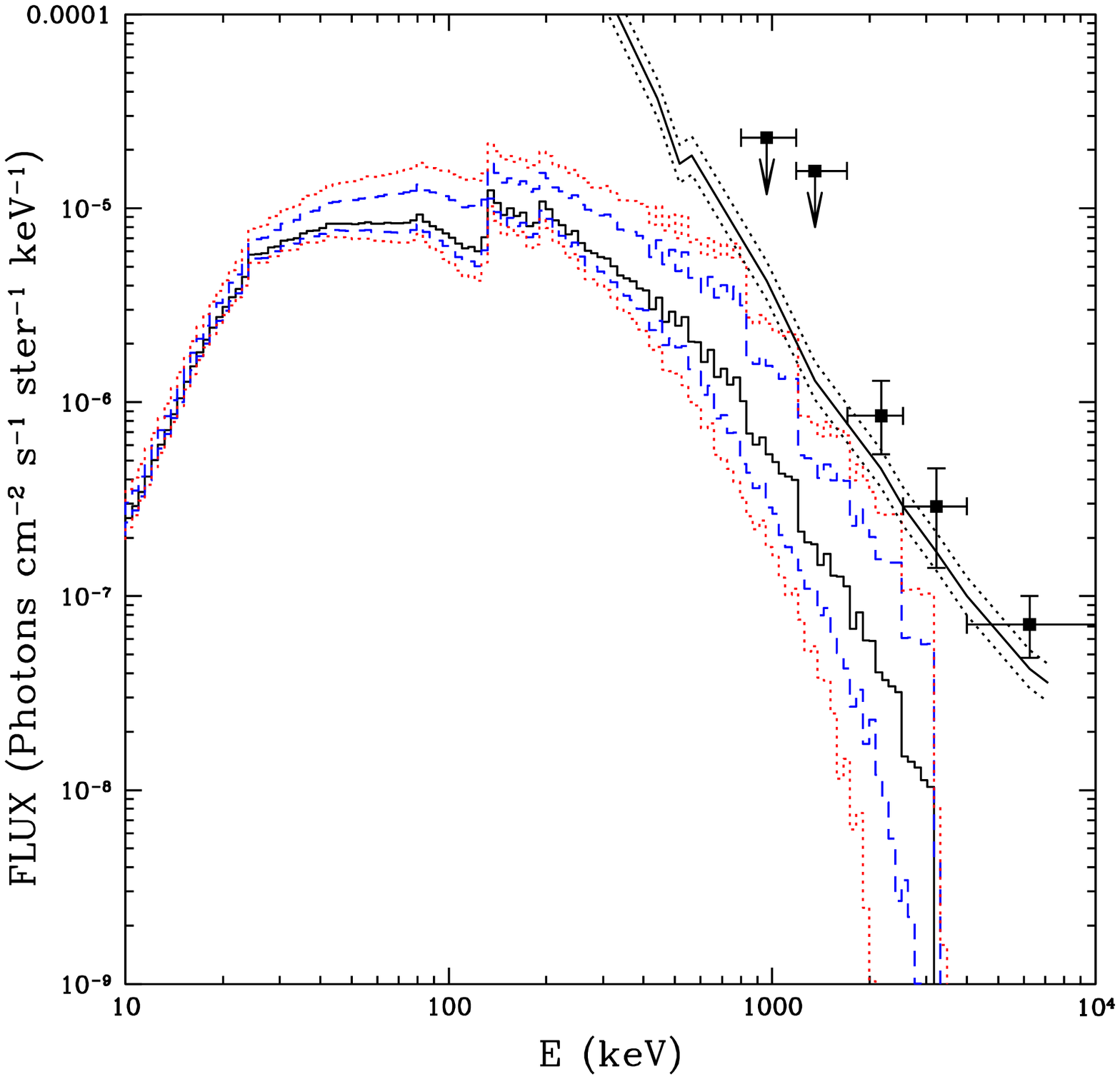}
\caption{
This figure shows the gamma--ray background using the Okumura et al. (2014) 
fit to the observed SNe Ia rates (black continuum line), together 
with its $\pm1\sigma$ bounds (blue dashed lines) and $\pm2\sigma$ bounds 
(red lines). 
The $\pm1\sigma$  and $\pm2\sigma$ bounds have been
traced by calculating the total error in the rates from the statistical and 
systematic errors of the points measured by these authors. The upper 
systematic errors include 50\% extinction by dust.   
The model used is the fully 
mixed W7fm (W99b), but predictions for the models W7, W7dt, N100, and the two 
models of the violent merging of a couple of WDs are qualitative similar.}
\label{Figure 9}
\end{figure}

In this work, for completeness, we calculate the background fluxes using the 
SNe Ia rates derived from the fit to the rates compiled by
 Graur et al. (2014) (whicch include their own CLASH data) (Fig 10) and
by Rodney et al. (2014) (CANDELS) (Fig. 11).
 The Rodney et al. (2014) rates, shown as the green shot--dashed line
  in Figure 5 and 6, fall 
much lower than Okumura et al. (2014) rates. And the corresponding 
gamma--ray background fluxes are lower than the fluxes
produced by using the rates from Okamura et al. (2014) (see Figure 10, 11). 
HB2010 (Fig 12) prediction is shown here for the purpose 
of checking for consistency when reproducing their estimate. This estimate,
based on more modest rates than those from Okumura et al. (2014) fit, 
would not give enough background to explain the MeV emission, either.

One might argue that the SN Ia rates toward higher redshifts could be 
underestimated if one does not take into account the distribution of 
SN Ia light curve stretches, because that would disfavor the detection of
 small--stretch, lower luminosity supernovae. However, all the groups 
mentioned above have taken due account of this effect in modeling observed rates.

\begin{figure}
\centering
\includegraphics[width=0.6\columnwidth]{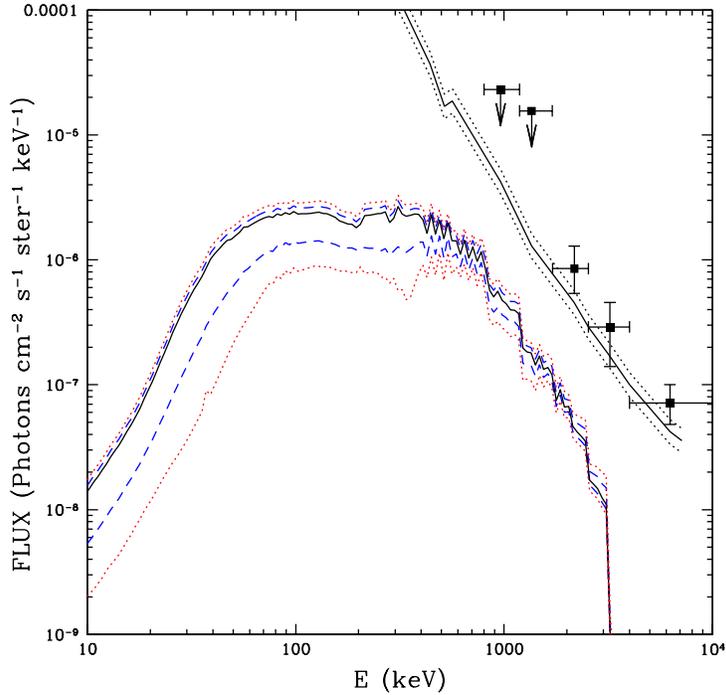}
\caption{ This figure shows the gamma--ray background using the Graur
et al. (2014) compilation of SNe Ia rates (see Figures 5 and 6)
as the solid line, and the upper 1$\sigma$ (blue 
dashed line) and 2$\sigma$ bounds (red dotted line). With those rates, the 
SNe Ia can not account for the observed gamma--ray background at the
2\,$\sigma$ level (nor even at the 3\,$\sigma$ level). The model used is the 
fully mixed W7 (W99b), but predictions for the other models are qualitatively 
similar.}
\label{Figure 10}
\end{figure}

\begin{figure}
\centering
\includegraphics[width=0.6\columnwidth]{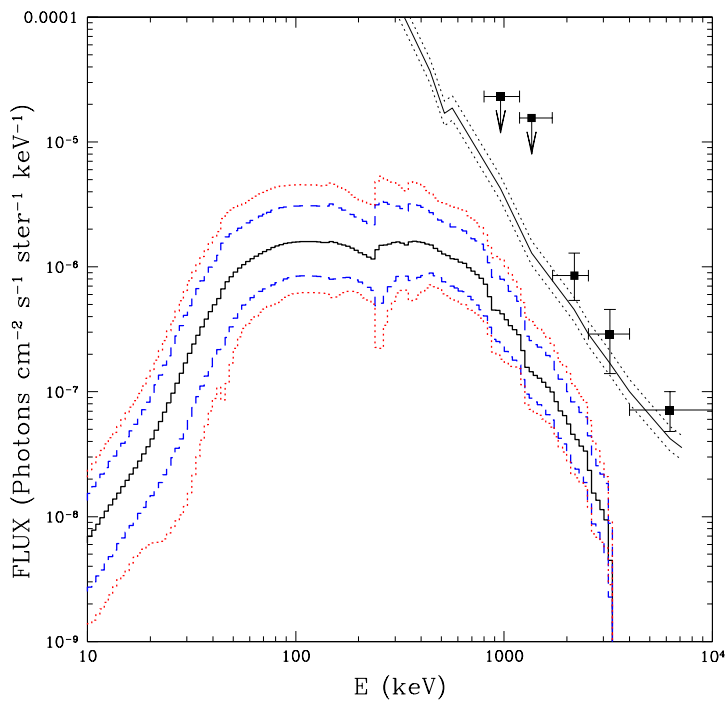}
\caption{This figure shows the gamma--ray background using the Rodney
et al. (2014) (CANDELS)  fit to the observational SNe Ia rates
with their prediction (solid black line) and the upper 1\,$\sigma$ (blue 
dashed line) and 2\,$\sigma$ bounds (red dotted line). With those rates, the 
SNe Ia can not account for the observed gamma--ray background at the
2\,$\sigma$ level (nor even at the 3\,$\sigma$ level). The model used is the 
fully mixed W7 (W99b), but predictions for the other models are qualitatively 
similar.}
\label{Figure 11}
\end{figure}

\begin{figure}
\centering
\includegraphics[width=0.6\columnwidth]{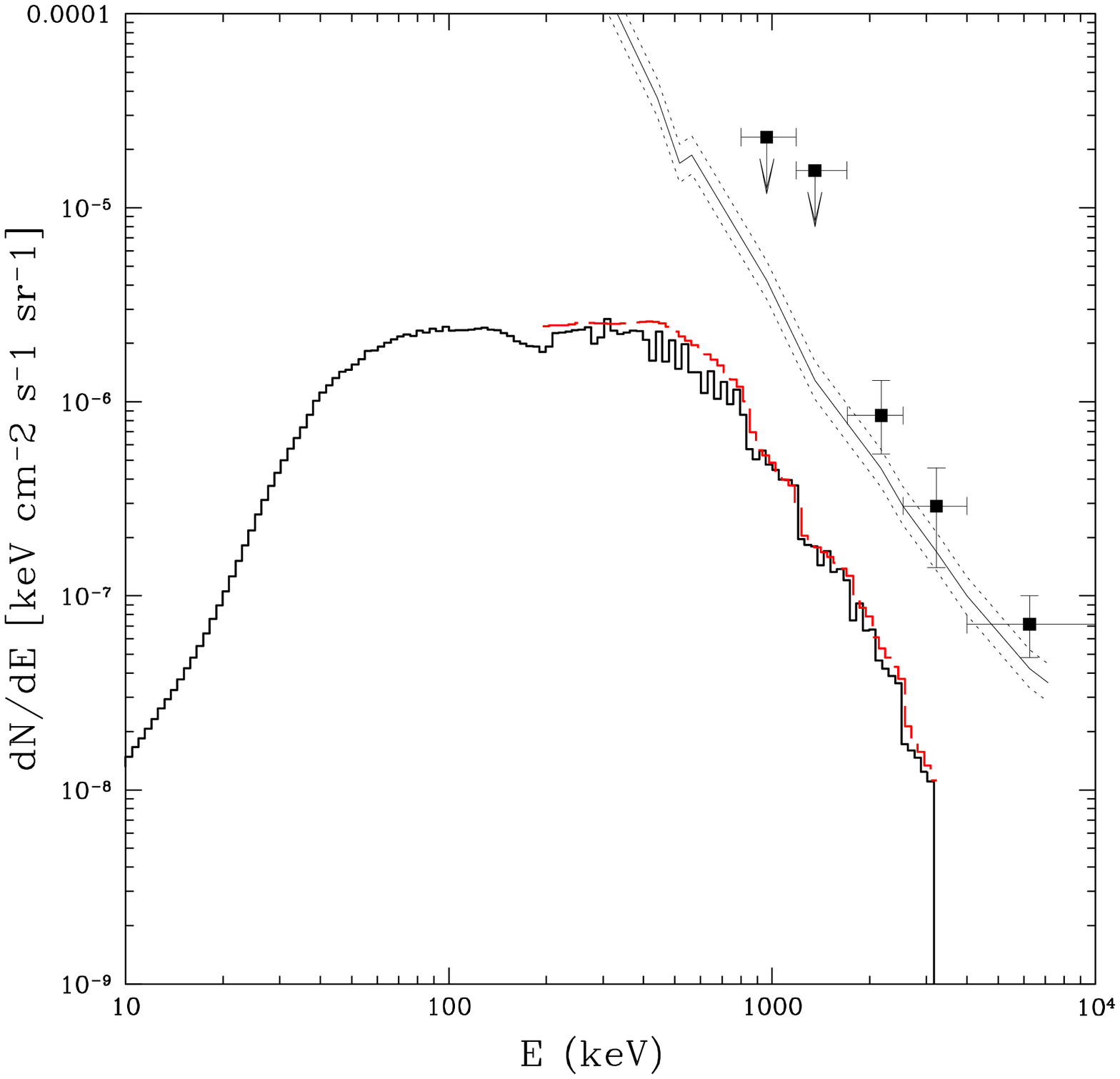}
\caption{ This figure shows the gamma--ray background using the HB2010
rates (private communication), together
with their prediction (solid red line). The model used is the 
fully mixed W7 (W99b), but predictions for the other models are qualitatively 
similar. We use in red the calculations provided by HB2010.
Our calculations using their rates  are shown in black and 
they agree with theirs.}
\label{Figure 12}
\end{figure}

\section{Discussion}

SNe Ia are known to be prominent contributors to the annihilation
line of positrons at 511 keV and to the positronium continuum 
observed by {\it OSSE} an by {\it INTEGRAL}. 
Milne et et al. (2001) estimate that 30--50 \%
of Galactic positrons may be explained by SNe Ia and massive stars.
The maximum of the contribution occurs at low bulge to disk ratios. 
Uncertainties of this contribution come from the SNe Ia Galactic rates
(Milne et al. 2001). Such non--fully attribution to SNe Ia  of the 
511 keV line has led to the suggestion that most of the contribution 
comes from annihilation of dark matter (Fayet 2004;  
Boehm et al. 2004). In the GeV regime, the lack of a full contribution 
from  astrophysical sources such as blazars, led to the  
proposal that weakly interacting massive particles 
(WIMPs) annhilations in the Galactic halo could be the main contributor to 
the diffuse GeV background fluxes. Those WIMPS would have masses 
in the range from 0.1 GeV to 10 GeV, for a variety of dark--matter halo 
models (Pullen, Chary, \& Kamionkowski 2007). However, recent work (Ajello 
et al. 2015; Di Mauro \& Donato 2015) greatly constrain the dark matter
contributions in this energy range. These authors find that blazars, together
with star--forming and radio galaxies, can account for the gamma--ray
background there.

\begin{figure}
\centering
\includegraphics[width=0.6\columnwidth]{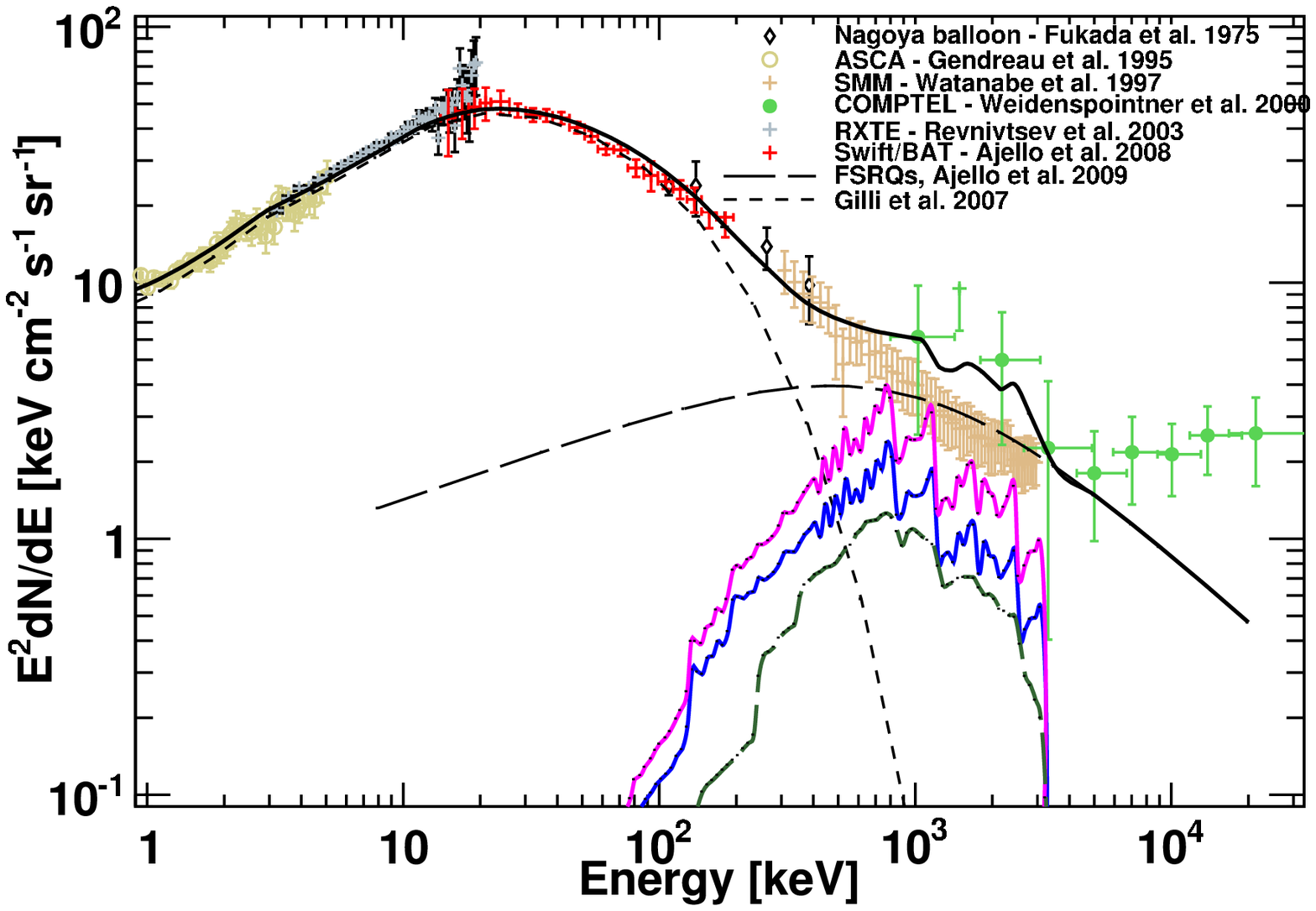}
\caption{The cosmic gamma--ray background in the 1 to 30,000 keV range, from 
different sources. Shown in the dash--dotted dark green line is the predicted
contribution by SNe Ia, when adopting the 2\,$\sigma$ upper limit of the SN Ia 
rates from {\it CANDELS} (Rodney et al. 2014), while in blue is that 
corresponding to the 1$\sigma$ upper bound of the rates by
(Okumura et al. 2014) and in magenta we show the gamma--ray background from the
 upper 2$\sigma$ to the rates from {\it Subaru/XMM--Newton} 
(Okumura et al. 2014), the magenta line (2$\sigma$ upper bound) being very 
unrealistic. The long--dashed black line is the contribution from 
FSRQs (Ajello et al. 2009), while the solid black line is the total gamma--ray 
background of the long--dashed black line (FSRQs) and background from the
 for 1$\sigma$ upper bound of SNe I rates
 from {\it Subaru/XMM--Newton} (Okumura et al. 2014). 
This gives a good fit to the observed gamma--ray background of SMM 
(Watanabe et al. 1997) but only because of the FSRQ contribution.} 
\label{Figure 13}
\end{figure}

In a similar way, the early estimates of SN Ia rates at high $z$ (D04), 
predicted a small contribution from Type Ia SNe to the gamma--ray 
background (AKH04). In that calculation, the SN Ia contribution is so small 
that even the 2\,$\sigma$ upper bound derived from the D04 SNe Ia rates 
cannot account for the measured gamma--ray background (AKH04, and Figure 7 of 
this work). This opened a window for contributions 
from WIMP decays in the MeV range (see, for instance, AKH05; 
Cembranos, Feng \& Strigari 2007), or from new populations 
of quasars (Ajello et al. 2009), amongst others. Here we show that the 
recent estimates of SNe Ia rates at high $z$, taken at face value, can not 
provide an explanation for the cosmic background in the MeV range.

Recently, Ajello et al. (2009) analyzing the three year data from the
Swift/Burst Alert Telescope survey concluded that the Flat--Spectrum
Radio Quasars (FSRQs) could account for most, or even all, of the cosmic gamma---ray 
background for energies above 0.5 MeV by assuming its inverse Compton peak 
is located in the MeV band. We show in Figure 13 as the long--dashed 
black line the best estimate of the contribution of FSRQs to the cosmic 
gamma--ray background. In Figure 13, the solid black line is the total
contribution of FSRQs and SNe Ia when we adopt the 1 sigma upper bound of
SNe Ia rates from Subaru/XMM--Newton (Okumura et al. 2014).
The decline of the FSRQs which match the obeservations 
above 10 MeV is dN/dE $\sim$ E$^{-2.5}$. 
The shape of the SNe Ia contribution is very similar up to 3 MeV. 
Clearly the gamma--ray background from SNe Ia can improve the fit 
of FSRQs to the measured gamma--ray background between 0.3 and 0.8 MeV.

\section{Conclusions}

Our calculations demonstrated that the use of SNe Ia rates "as measured" at 
various  redshifts are  too low to account for the bulk of the observed 
gamma--ray background in the MeV range. We find that, above $z$ = 2--2.5, the 
gamma--ray emission from  SNe Ia makes no significant contribution to the cosmic  
background. It is thus not expected  that future determinations of the SNe Ia 
rates above that range would make a noticeable difference. However, what does
make a difference is the overall SNe Ia rate from $z = 0$ to $z = 2$. We have shown 
here that recently reported values of the SNe Ia rates from Okumura et al. (2014), 
Graur et al. (2014) and Rodney et al. (2014) predict a contribution that is about a factor 
of five below the observed background. Contributions from $z$ beyond 2 are, as noted, 
too distant to be significant, even in the most optimistic cases, where rates continue to 
grow (see Figure 5 and 6). 

A relevant aspect of the uncertainties in determining rates at high $z$ ($  >  1.4$) is that 
the detection efficiency  decreases rapidly with redshift, as observed bands shift 
farther into the rest--frame UV spectrum, where SN Ia emission is scarce 
(see the spectra shown in  Riess et al. 2007). This causes the uncertainties  to 
become very large at z $\sim$ 1.6 (Barbary et al. 2012). This question and the one
of addressing dust obscuration which requires a model for cosmic dust evolution,
makes the measurements of the rates at high z very unreliable. Note, however, the good 
agreement between the plethora of measurements of SNe Ia at low z (Figure 6). 

The current measurements of rates at high $z$ concentrate on normal 
(cosmological) SNe Ia. They make up to about 60\% of all
SN Ia explosions (see Ruiz--Lapuente 2014, for a review). The average 
predictions for the gamma--ray background, which at face values give
gamma--ray backgrounds that are too low, might slightly change if the bulk 
of the non--cosmological SNe Ia were included. It has been estimated that
up to 30\% of SNe Ia could be of the Type Iax (Foley et al. 2013), though a
lower estimate is found in White et al. (2014).  
Those SNe Ia produce smaller amounts of $^{56}$Ni, thus making an even 
smaller contribution to the background (although their smaller ejecta mass 
increases the gamma--ray escape fraction). Furthermore, there is also a 15\% group
of SN1991bg--like events, which are much less luminous than the cosmological SNe Ia. 

From the opposite perspective, a factor that could make the mean 
gamma--ray fluxes larger would be to adopt models brighter in gamma--rays 
than W7 or W7fm (both models giving about the same average contribution over 
600 days after explosion); some SNe Ia models can give twice as many escaping 
gamma--rays. One should acknowledge, however, that those models might 
fare worse in the comparison with observed spectra and light curves in the 
optical range. The first SN Ia observed in gamma--rays, SN 2014J 
(Churazov et al. 2014; Diehl et al. 2014), can be well reproduced within the 
set of realistic models presented in this paper.  SN 2014J synthesized 
an amount of $^{56}$Ni that falls within the standard range (about 0.5--0.6 M$_{\odot}$). 
That supernova is best explained as the explosion of a white dwarf close
to the Chandrasekhar mass (Diehl et al. 2014). Having considered a wide range of models, 
we find that this is not the decisive point in regards to the question we
address in the paper, i.e., whether SNe Ia are major contributors to
the gamma--ray background in the MeV range. 

In summary, following many measurements and estimates of the SNe Ia rate
as a function of redshift, we now have a more reliable estimate of the SNe Ia induced 
gamma--ray background in the MeV range than it was possible in the last XXth century and beginning of the XXIst. As the yields of different SN Ia models 
do not vary greatly, the conclusions are nearly independent of the particular model taken 
to represent normal SNe Ia. Attributing the gamma--ray background to SNe Ia appears 
tempting, as we know that they are significant $\gamma$--ray sources, but within the current 
uncertainties SNe Ia do not seem able to account for all of the MeV background. One  
needs to look for another source of the  gamma--ray background in the MeV 
range, with FSQRs as a promising object class. SNe Ia can produce up to 30--50$\%$ 
of the gamma--ray background emission (if we take observed event rates at face value) and 
the remaining half to two thirds would have to be attributed to one or more contributors.

\section{Acknowledgements}
PLR would like to thank the Max--Planck--Institut f\"ur Astrophysik, where 
this project was started.  
LST would like to thank the Palmetto Cluster administrative and support team 
of Clemson University for giving us the computational time to finish this work.
The authors do also thank Jun Okumura and Steve Rodney for useful information
concerning the observed supernova rates. We thank as well Shunsaku
 Horiuchi and John
 Beacom for providing their predictions for rates and background.
 Our thanks go as well to 
Ivo Seitenzahl and R{\"u}diger Pakmor for providing data of their
 Type Ia supernova explosion models.
PRL and RC acknowledge support from grant AYA2012--36353, of the Ministerio
de Econom\'{\i}a y Competitividad (MINECO) of Spain. 
FKR receives support from the Emmy Noether Program (RO 3676/1--1) of DFG and 
the ARCHES prize of the German Federal Ministry of Education and Research 
(BMBF). 
STO acknowledges financial support from Studienstiftung des deutschen
Volkes.

\end{document}